\documentclass[9pt,conference]{IEEEtran}
\usepackage{graphicx}
\usepackage{epsfig}
\usepackage{amssymb,amsfonts}
\usepackage{amsmath}
\usepackage{setspace}
\usepackage{lscape}
\usepackage{spconf}

\newcommand{\Real}{\mathbb{R}}

\def \bS{\mathbf{S}}

\def \bU{\mathbf{U}}
\def \bX{\mathbf{X}}
\def \bY{\mathbf{Y}}
\def \bV{\mathbf{V}}

\def \bL{\mathbf{L}}
\def \bY{\mathbf{Y}}

\def \b1 {\mathbf{1}}

\begin{document}

\title{Exploiting structural complexity for robust and rapid hyperspectral imaging}
\name{Gregory~Ely, Shuchin~Aeron and Eric~L.~Miller \thanks{ Will need to add NSF GRFP bit here.}}
\address{Department of ECE, 161 College Avenue, Tufts University, Medford, MA, 02155}
\thanks{This material is based upon work supported by the National Science Foundation Graduate Research Fellowship under Grant No. ???}

\maketitle
\begin{abstract}
This paper presents several strategies for spectral de-noising of hyperspectral images and hypercube reconstruction from a limited number of tomographic measurements. In particular we show that the non-noisy spectral data, when stacked across the spectral dimension, exhibits low-rank. On the other hand, under the same representation, the spectral noise exhibits a banded structure. Motivated by this we show that the de-noised spectral data and the unknown spectral noise and the respective bands can be simultaneously estimated through the use of a low-rank and simultaneous sparse minimization operation without prior knowledge of the noisy bands. This result is novel for for hyperspectral imaging applications. In addition, we show that imaging for the Computed Tomography Imaging Systems (CTIS) can be improved under limited angle tomography by using low-rank penalization. For both of these cases we exploit the recent results in the theory of low-rank matrix completion using nuclear norm minimization.
\end{abstract}
\begin{keywords}
Hyperspectral imaging, de-noising, Limited angle tomography, low-rank recovery. \vspace{-1mm}
\end{keywords}
\section{Introduction}
\label{sect:intro}

This paper addresses two specific image reconstruction challenges encountered in the field of hyperspectral imaging: de-noising in the presence of spectral noise and hypercube reconstruction from a limited set of Radon projections similar to angle limited Computed Tomography Imaging Systems (CTIS).

The first of these two problems is motivated by the desire to remove noise at specific frequency bands from hyperspectral image cubes.  This problem frequently arises when using satellites or aircraft to capture hyperspectral images of the earth in which the light reflecting from the surface of the earth must travel through several kilometers of atmosphere to the sensor.  The atmosphere even without the presence of clouds has extremely high absorption bands, particularly at 1400 nm and  1900 nm due to water in the atmosphere \cite{aviris}.  This effect leads to numerous bands being discarded for many data classification and analysis algorithms \cite{kaewpijit_wavelet-based_2002} \cite{shah_recent_2004}.

In order to mitigate the effects of both this spectral and electronic noise several de-noising techniques such as multidimensional Weiner filtering \cite{letexier_multidimensional_2008} and methods exploiting the use of high order singular value decompostion \cite{chen_denoising_2011}, curvelets \cite{lei_junk_2011}, and wavelets \cite{kaewpijit_wavelet-based_2002} \cite{scheunders_least-squares_2004} have been used to de-noise these effects.  However, both the intensity dependence the noise and the concentration across a few spectral bands makes the removal of optical noise challenging \cite{acito_signal-dependent_2011}. Many of these techniques are based on the premise of noise being AWGNG and performance can be poor \cite{liu_nonwhite_2012}.  Typically a preprocessing (whitening) step is needed to mitigate the effects of the Poisson noise and improve performance \cite{liu_nonwhite_2012}.  Recently,  efforts to de-noise spectral bands have focussed on the use of sparse or joint penalizations in an appropriate basis such as wavelets \cite{zelinski_denoising_2006} and dictionary learning techniques \cite{xing_dictionary_2012}.

In this paper we will explore a novel spectral de-noising technique based on a low-rank and simultaneously sparse matrix decomposition. The low-rank sparse matrix decomposition or Robust Principle Component Analysis (RPCA) has been well studied and theoretical limits well characterized in recent years \cite{candes_robust_2011} \cite{hsu_robust_2011}.  Furthermore, RPCA has been successfully employed in image and video processing to separate background from the foreground \cite{pope_real-time_2011} and remove `salt and pepper' noise from imagery \cite{candes_robust_2011}.  However, little research has been done to explore variations of RPCA such as a low-rank group sparse decomposition and its potential applications.  In particular, Tang proposed a feasible solution to solving the group RPCA problem through the method of Augmented Lagrange Multipliers \cite{tang_robust_2011} and Ji demonstrated the use of group RPCA to de-noise video data \cite{ji_robust_2011}. This paper provides another potential application and extension of RPCA to CTIS systems.

In the second part of this paper we will focus on the problem of estimation of the hyperspectral data cube from limited number of tomographic projections. Here we show how the use of low-rank regularization can be used to improve an existing class of hyperspectral imagers. These hyperspectral imagers \cite{descour_demonstration_1998,wagadarikar_spectral_2008,mooney_high-throughput_1997} sample the hyperspectral image cube by \emph{simultaneously} (i.e.  not sequentially) taking a number of Radon type projections of the 3D data cube onto a 2D focal plane array using diffractive optics. Traditionally filtered back-projection methods have been employed to recover the data cube form these tomographic projections.  However, these techniques need a large number of projections to ensure accurate results, avoid the so called missing cone problem \cite{Feng1996} and often fail in noisy environments. This need for a large number of projections increases the necessary focal plane size beyond what is often feasible.  In this context we demonstrate how one can exploit the low-rank regularization to improve the reconstruction under these classes of simultaneous and \emph{compressive measurements}.  Note that although some research has focussed on the use of both sparse and low-rank reconstructions of hyperspectral image cubes, these studies use \emph{practically infeasible} sampling techniques such as randomly sampling a small set of pixels within the image cube \cite{waters_sparcs:_2011}.

\begin{figure}[h]
\centering \makebox[0in]{
    \begin{tabular}{c}
      \includegraphics[height = 2 in, width = 3.7in]{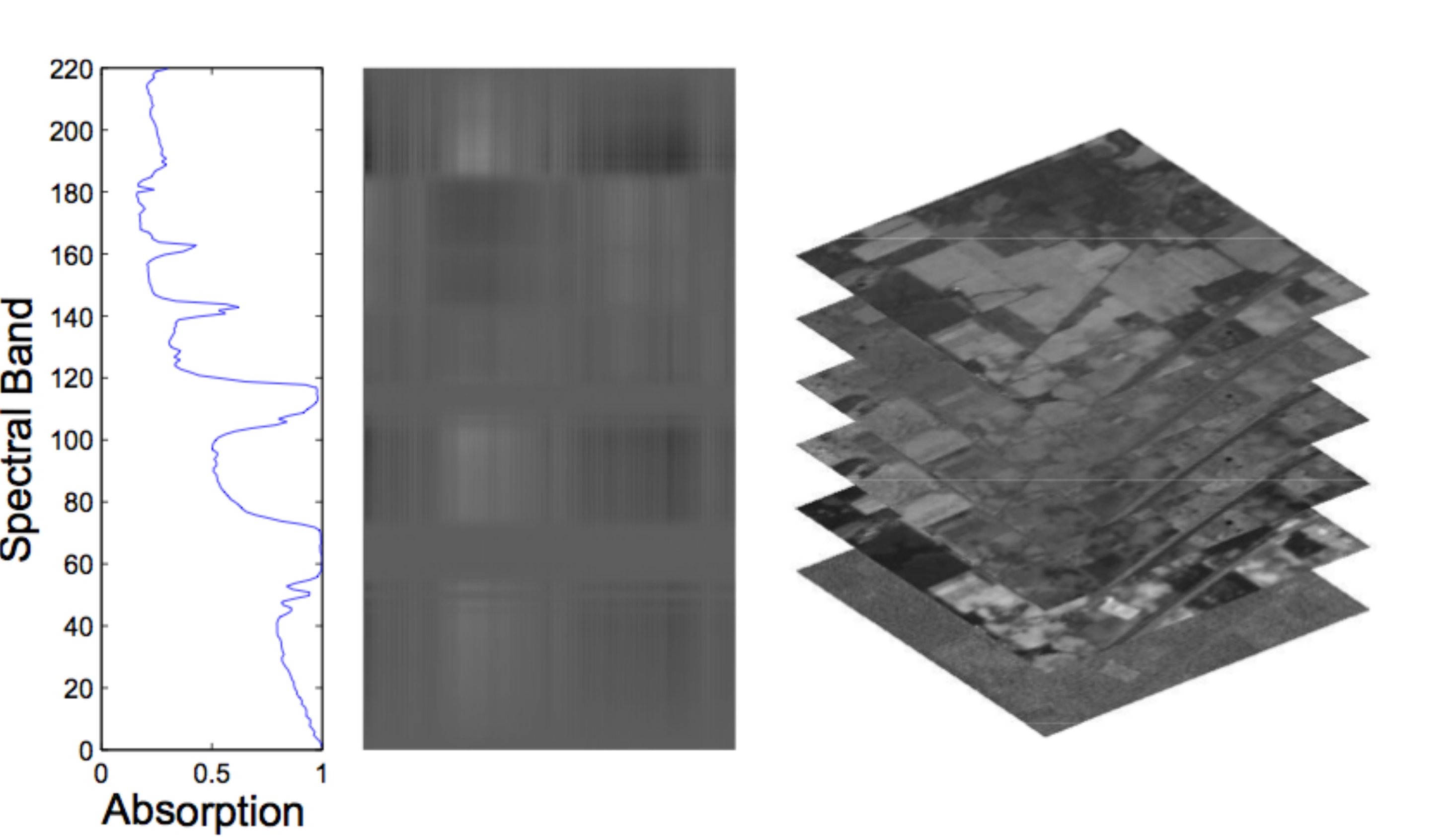}
      \end{tabular}}
  \caption{Left: Normalized total counts in the AVIRIS image as a function of band.  We see two pronounced absorption bands. Right \& Center: This figure shows a 3D and 2D representation of a hyperspectral image. One notices the horizontal bands of spectral noise in the two dimensional image that align with the absorption bands.  Much of the structure in the matrix appears to be vertical but the horizontal bands are spectral noise at absorption bands.}
  \label{fig:spectrumEmperical}
  \label{fig:threeTwoArray}
\end{figure}

%

\section{Structural complexity of hyperspectral images}
\label{sect:background}

A hyperspectral image or data cube consists of many images of the same size collected over a number of spectral bands.  Mathematically the hyperspectral image can be considered as a three-dimensional matrix $\bL \in \Real^{m\times n\times l}$ with spatial dimensions of $m$ and $n$ pixels and at $l$ wavelengths.  One can reshape  the hyperspectral image as a two-dimensional array with a number of columns equal to the number of spectral bands and where each column is the vectored image at the given band, see Figure~\ref{fig:threeTwoArray}. With slight abuse of notation we denote the reshaped image by $\bL \in \Real^{mn\times l}$. We now present two observations regarding the structural complexity of the image data which will be exploited for recovery and de-noising.

\vspace{-2mm}

\subsection{Low-rank structure of the hyperspectral data cube}
Although a hyperspectral image/data may have numerous bands, it has been shown that signal subspace is significantly smaller than the number of bands \cite{harsanyi_hyperspectral_1994} \cite{landgrebe_hyperspectral_2002}.   In particular the eigenvalues of the reshaped hyperspectral cube $\mathbf{L}$ obey a power law decay. This means the vector of eigenvalues has a small weak-$\ell_p$ norm \cite{Donoho_Uncond_ACH93} which implies that image is compressible under the suitable transformation.  This intuition can be physically explained by considering the Singular Value Decomposition (SVD) of the (reshaped) hyperspectral matrix $\bL$.
\begin{align}
    \bL = \bU \Sigma \bV^{*}
    \label{eq:svdBasic}
\end{align}
We can think of the right singular vectors as giving the spectra of the common elements in the scene and the left singular values as the concentration map of these spectra.  The singular values then give the relative amount each compound in the scene. Low-rank of the image can then be interpreted as presence of a few spectra with a correlated concentration profile across space.
\vspace{-2mm}
\subsection{Sparsity structure of hyperspectral noise}
In hyperspectral imaging the atmosphere can lead to vastly different absorption rates across the spectrum of interest.  In particular as shown in Figure~\ref{fig:spectrumEmperical}, the two water absorption bands are attenuated, roughly at band 60 and 100. In a typical hyperspectral data processing the data from these two bands would be discarded. On the other hand we note that in the noisy reshaped image, the spectral noise exhibits a banded structure which is mathematically equivalent to saying that hyperspectral noise exhibits a simultaneous sparse structure under the given reshaping of the data cube.


\noindent Therefore, the noisy reshaped hyperspectral data cube can be represented as $\bY = \bL + \bS$ where $\bL$ is the low-rank non-noisy image and $\bS$ is the spectral noise which is simultaneously or group sparse across bands.
\vspace{-2mm}
\section{Robust \& rapid hyperspectral imaging}
\label{sect:mathFrame}
Both the spectral de-noising and limited angle reconstruction problems can be viewed through the following framework in which we observe noisy measurement, $\bY$ of hyperspectral image cube $\bL$ through a measurement system described by the linear operator (matrix) ${\cal A}$, i.e.
\begin{align}
\label{eq:thingy11}
\bY = {\cal A} (\bL + \bS)
\end{align}
The problem is that given the observation $\bY$ and the \emph{sensing} operator ${\cal A}$ (to be defined below for both problems of interest) we want to recover the de-noised image $\bL$ while removing the noise $\bS$.
\vspace{-1mm}
\subsection{Complexity penalized recovery algorithms}
\vspace{-1mm}
To de-noise and recover the hyperspectral data, one can exploit the low-rank and sparse structure of the data and noise and solve the following optimization,
\begin{align}
\label{eq:theMinEquation}
\min_{\textbf{L},\bS} ||{\cal A}(\textbf{L}+\bS)(:) - \bY||_2^2 + \lambda_L \textrm{rank(\textbf{L})} + \lambda_S ||\bS||_{0,2}
\end{align}
where $\lambda_L$ \& $\lambda_S$ control the relative strength of the sparsity and low-rank penalization and $|| \bS||_{p,q} $ is the $p$-norm of the vector formed by taking the $q$-norm along the rows of $\bS$ or otherwise also known as $\ell_{p,q}$ norm. This optimization problem is known to be NP-hard. However, the rank and support penalties can be relaxed to the nuclear norm and $\ell_{1,2}$  norm, respectively, which makes the optimization tractable while  still encouraging the desired structure for $\bL$ and $\bS$ \cite{candes_robust_2011}. Therefore we relax the above combinatorial optimization problem to the following convex optimization problem and
consider three cases.
\begin{align}
\label{eq:minRelaxed}
{\min_{\textbf{L},\bS} ||{\cal A}(\textbf{L}+\bS) - y||_2^2 + \lambda_L ||\textbf{L}||_* + \lambda_S} ||\bS||_{1,2}
\end{align}
\noindent \textbf{Case I - Hyperspectral de-noising with raster scan data} -
In this case ${\cal A}$ is an identity operator and therefore the optimization problem becomes,
\begin{align}
\label{eq:spectralNoise}
\min_{\bL,\bS} || \bL +\bS - \bY||_2^2 + \lambda_L ||\bL||_* + \lambda_S ||\bS||_{1,2}
\end{align}

In Section~\ref{sect:results} we will demonstrate the performance of this algorithm on real hyperspectral data and give experimental results that motivate why the sparse component is necessary for the de-noising.\\
\noindent\textbf{Case II. - Image recovery from limited angle tomography: No spectral noise} -
As pointed out in the introduction the CTIS systems are limited by the size of focal plane array which limits the number of tomographic projections that can be obtained.  In this case traditional reconstructions suffer from the missing cone problem \cite{Feng1996}. These methods however do not exploit the low complexity of the underlying data cube. Assuming no spectral noise, given the limited number of Radon projections we propose the following algorithm for estimation of the hyperspectral image which exploits the low-rank structure.
\begin{align}
\label{eq:radon}
\min_{\bX} ||{\cal A}(\bX) - \bY ||_2 + \lambda ||\bX||_{*}
\end{align}

\noindent\textbf{Case III. - Image recovery from limited angle tomography: Noisy case} -
Here we consider the most general case where the spectral data is corrupted by banded spectral noise and the data is acquired through a CTIS system with limited number of Radon projections. In this case
simultaneous spectral cube recovery and spectral de-noising is affected by solving for the optimization problem given by equation \ref{eq:minRelaxed}. In the next section we will present detailed experimental results of the proposed algorithms on real data sets.

\section{Experimental evaluation}
\label{sect:results}
\begin{figure}[h]
\centering \makebox[0in]{
    \begin{tabular}{c}
      \includegraphics[height = 2.5in, width = 4in]{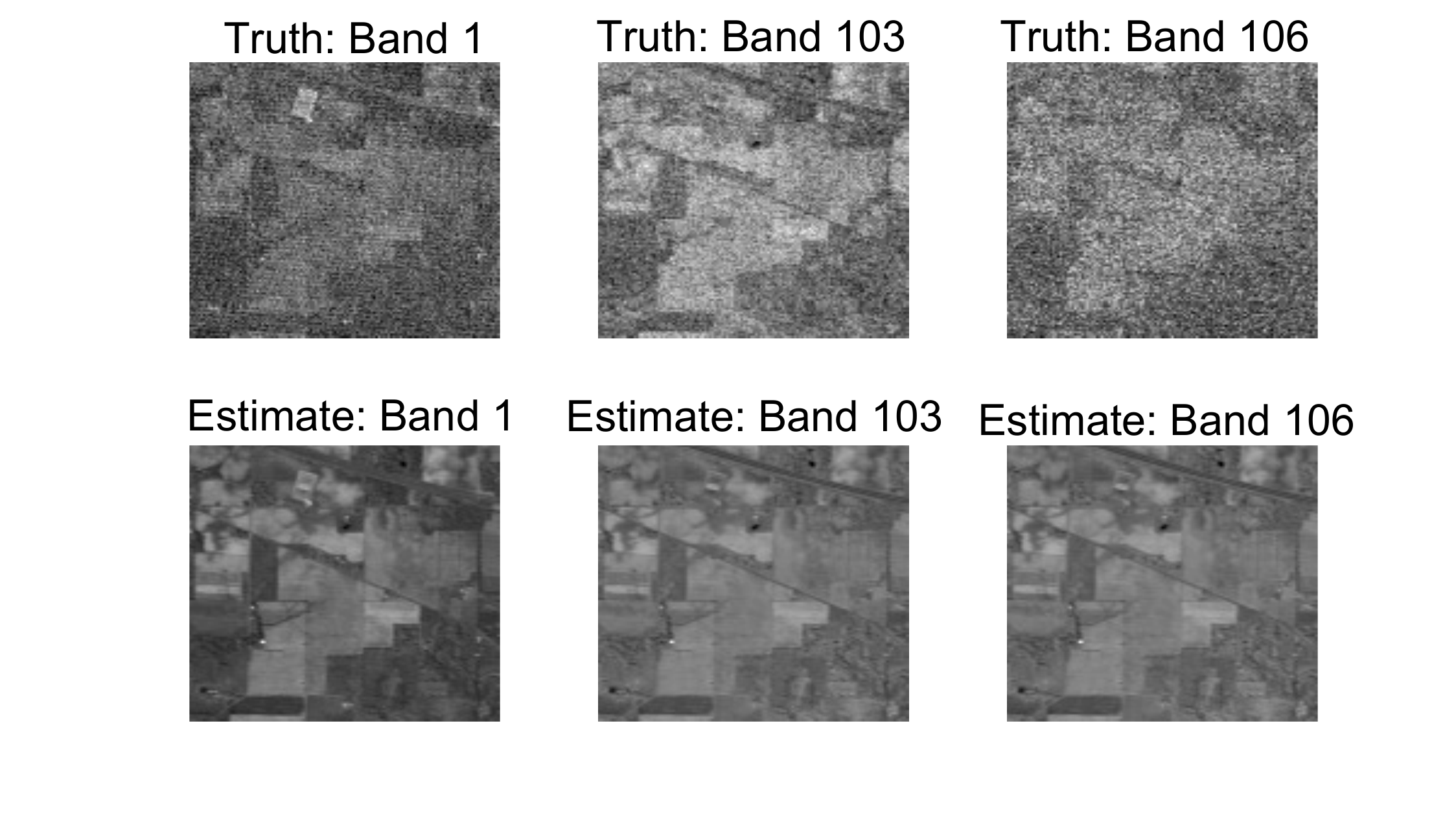}
      \end{tabular}}
  \caption{This figure shows images from AVIRIS data at various bands before de-noising and after de-noising.}
  \label{fig:DenoisedBands}
\end{figure}

In this section we will use a real hyperspectral image taken from Airborne Visible/Infrared Imaging Spectrometer (AVIRIS) [website- http://aviris.jpl.nasa.gov/html/aviris.spectrum.html], far above an rural scene with a spatial dimension of 128 by 128 pixels. The imager uses 220 bands which cover the spectrum from IR  to visible range. The two water absorption bands centered at 1400 and 1900 nm corrupt the image.  \textbf{NOTE}: All the optimization problems below are implemented using TFOCS \cite{tfocs}.\\
\noindent \textbf{Case I. - Hyperspectral de-noising} -
\label{sect:hyperDenoise}
In some of the less noisy bands the structure of the image is still somewhat visible (figure \ref{fig:DenoisedBands}).  In order to improve the de-noising the AVIRIS data we first take and record the Frobenious norm of each frame to construct a $N_\lambda \times 1$ vector $W$.  We then use this recorded vector to normalize each image at given wavelength such that the signal energy in each band is 1.   Because we expect the noise in our experiment to be due to low photon counts in bands of high absorption, we can use the vector $W$ to weight the minimization operation. In particular we want to encourage row sparsity along the bands with low counts.  In order to do so we modify  equation \ref{eq:spectralNoise} to include the weighting factor $W$, that makes it more expensive for the intense bands to be decomposed into the sparse matrix.
\begin{align}
\label{eq:spectralNoiseMod}
\min_{\bL,\bS} ||(\bL+\bS) - \bY||_2^2 + \lambda_L ||\bL||_* + \lambda_S ||W\bS||_{1,2}
\end{align}
This weighting factor allows our algorithm to be more robust to choices of $\lambda_S$ and $\lambda_L$ as it effectively decreases the coherence between the $\ell_{1,2}$ norm and the nuclear norm.

\begin{figure}[h]
\centering \makebox[0in]{
    \begin{tabular}{c}
      \includegraphics[height = 1.25in, width = 3.75in]{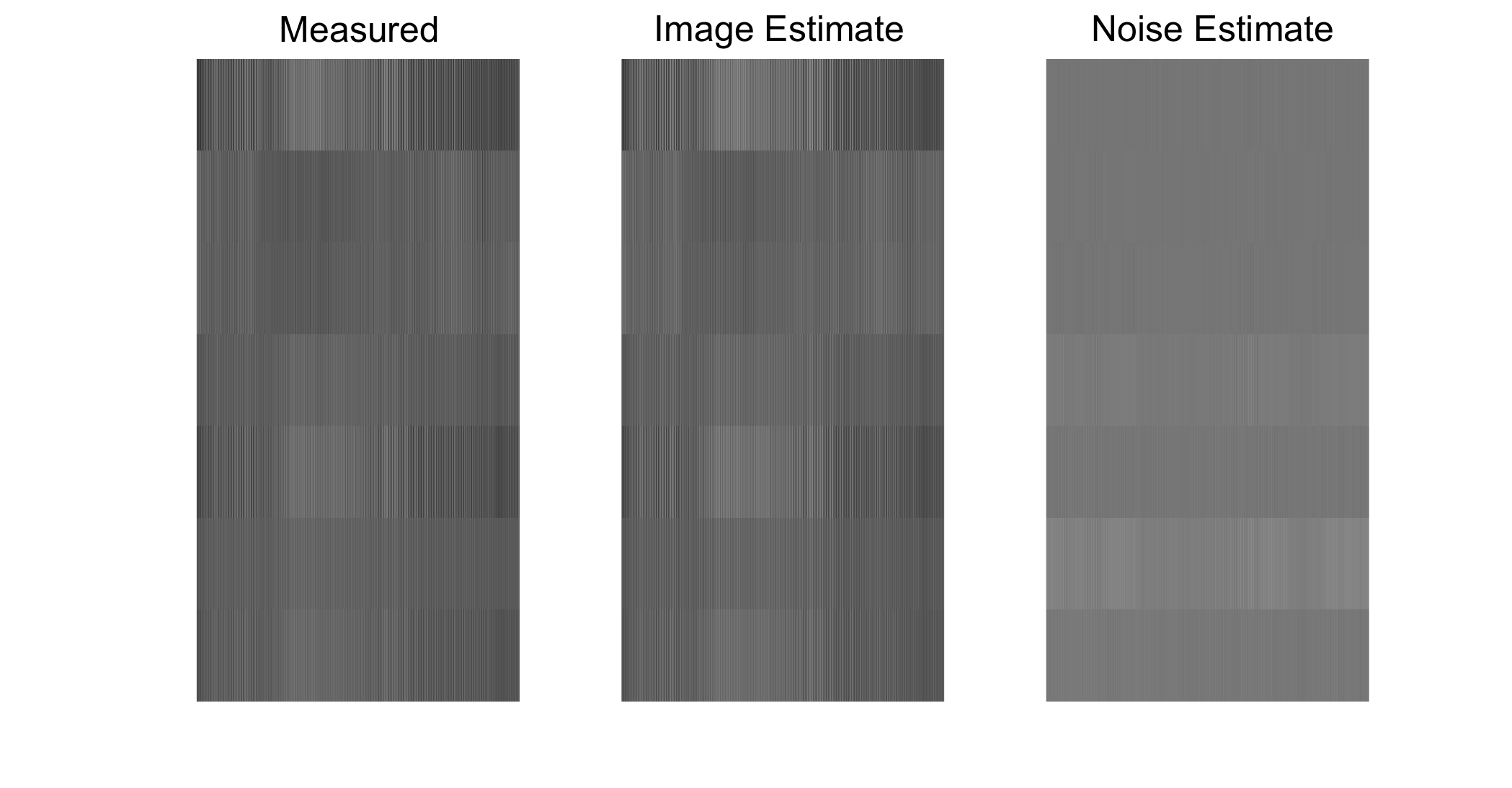}
      \end{tabular}}
  \caption{This figure shows 2D hyperspectral cube with noise and low-rank reconstruction.}
  \label{fig:twoD}
\end{figure}

The minimization operation in Equation (\ref{eq:spectralNoiseMod}) is then applied to the hyperspectral image with $\lambda_S$ of .06  and $\lambda_L$ of $0.1$. The proposed algorithm was successful in de-noising and was able to remove the spectral noise.  Figure \ref{fig:DenoisedBands} show the results of algorithm applied to a few very noisy bands and figure \ref{fig:twoD} shows the results applied to all bands of the hyperspectral image.  Like in the synthetic example, we can now see structure in the bands that were previously noisy.

\begin{figure}[h]
\centering \makebox[0in]{
    \begin{tabular}{c}
      \includegraphics[height = 1.75in,  width = 3.5in]{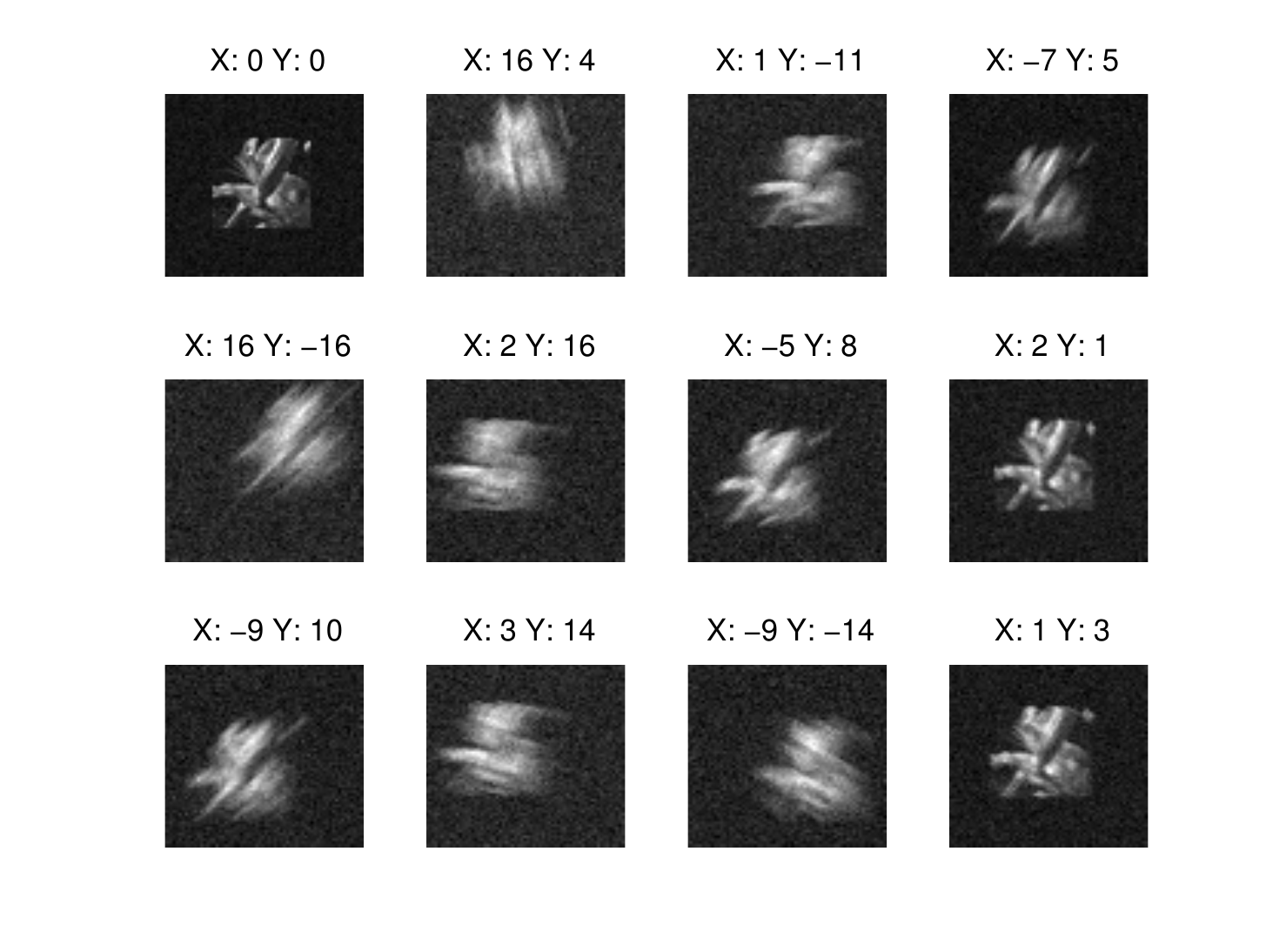}
      \end{tabular}}
  \caption{This figure shows the 12 noisy radon projections of the hyperspectral image cube.  With 12 projections the system is underdetermined.}
  \label{fig:radonProj}
\end{figure}

\begin{figure}[h]
\centering \makebox[0in]{
    \begin{tabular}{c}
      \includegraphics[height = 1.2 in, width = 3.75in]{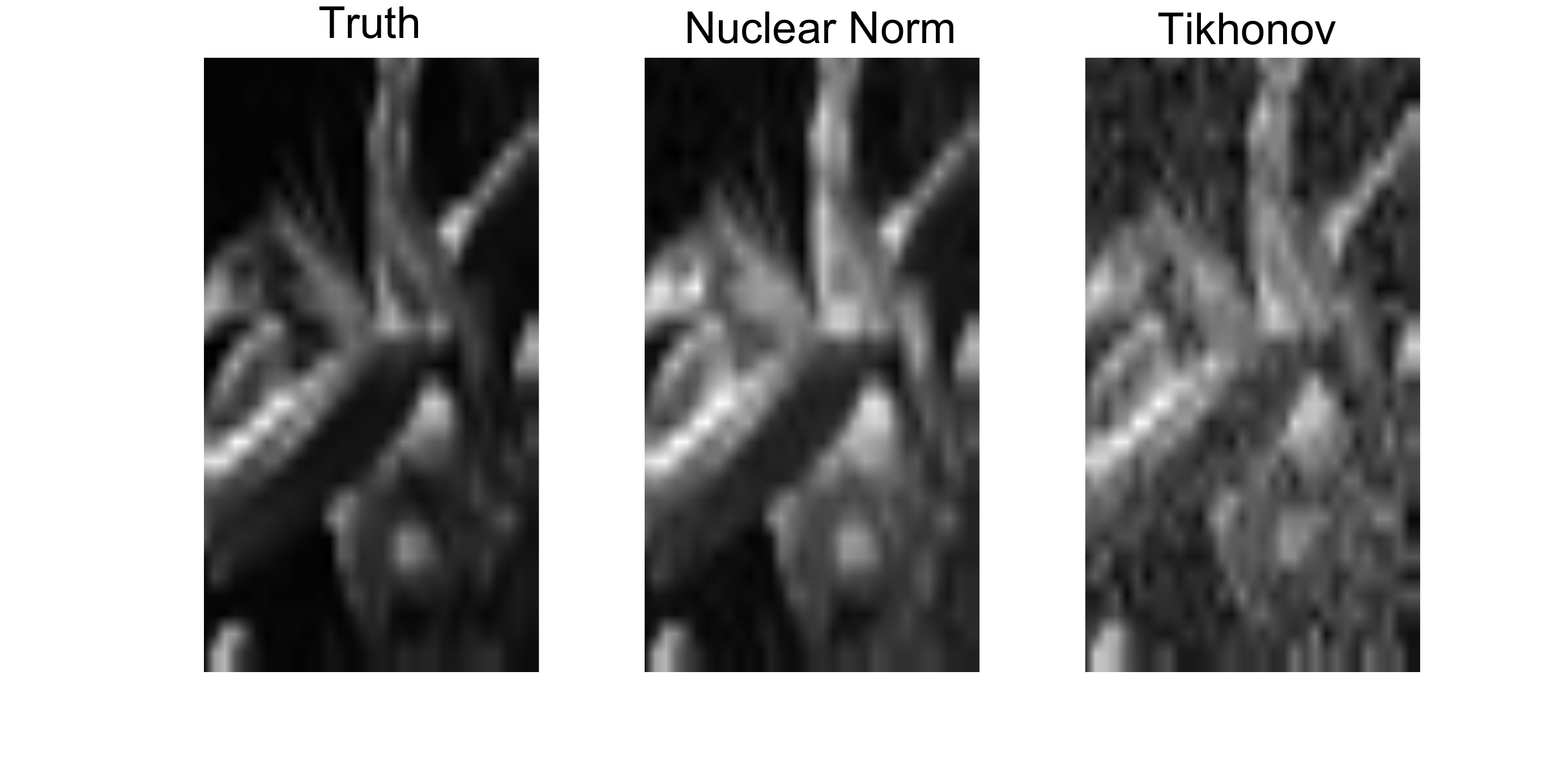}
      \end{tabular}}
  \caption{This figure shows an example of the true image, low-rank reconstruction, and least square reconstruction, from the hyperspectral flower at band 12.}
  \label{fig:radonFlower}
\end{figure}

\noindent \textbf{CaseII. - Hyperspectral imaging from limited Radon projections with no spectral noise} -
\label{sect:radon}
In the following example we attempt to reconstruct the 32 by 32 image of the hyperspectral flower \cite{foster_information_2004} using a limited number of projections. The projections at various angles for a typical single-shot CTIS system \cite{descour_demonstration_1998,johnson_spatial-spectral_2006} are shown in Figure~\ref{fig:radonProj}.  Gaussian noise was then added to the measured projections, such that the resulting SNR of the projections was 4.5 dB.  This projection operation can be represented through the underdetermined matrix ${\cal A}$.  In this case we solve the optimization problem (\ref{eq:radon}) for recovery.  We compare the performance of this method to the standard Tikhonov regularization approach with $\ell_2$ norm penalty instead of nuclear norm penalty.  The choice of $\lambda$ for both cases were determined using the Kolmogorov-Smirnoff (KS)-test method described in Section \ref{sect:param}.  As expected the low-rank minimization resulted in a better reconstruction of the hyperspectral image cube with normalized mean square error of .23 versus .35 for the Tikhonov reconstruction. The resulting reconstruction for the $12^{th}$ band is shown in Figure~\ref{fig:radonFlower}.

\begin{figure}[h]
\centering \makebox[0in]{
    \begin{tabular}{c}
      \includegraphics[height = 1.7 in, width = 3.5in]{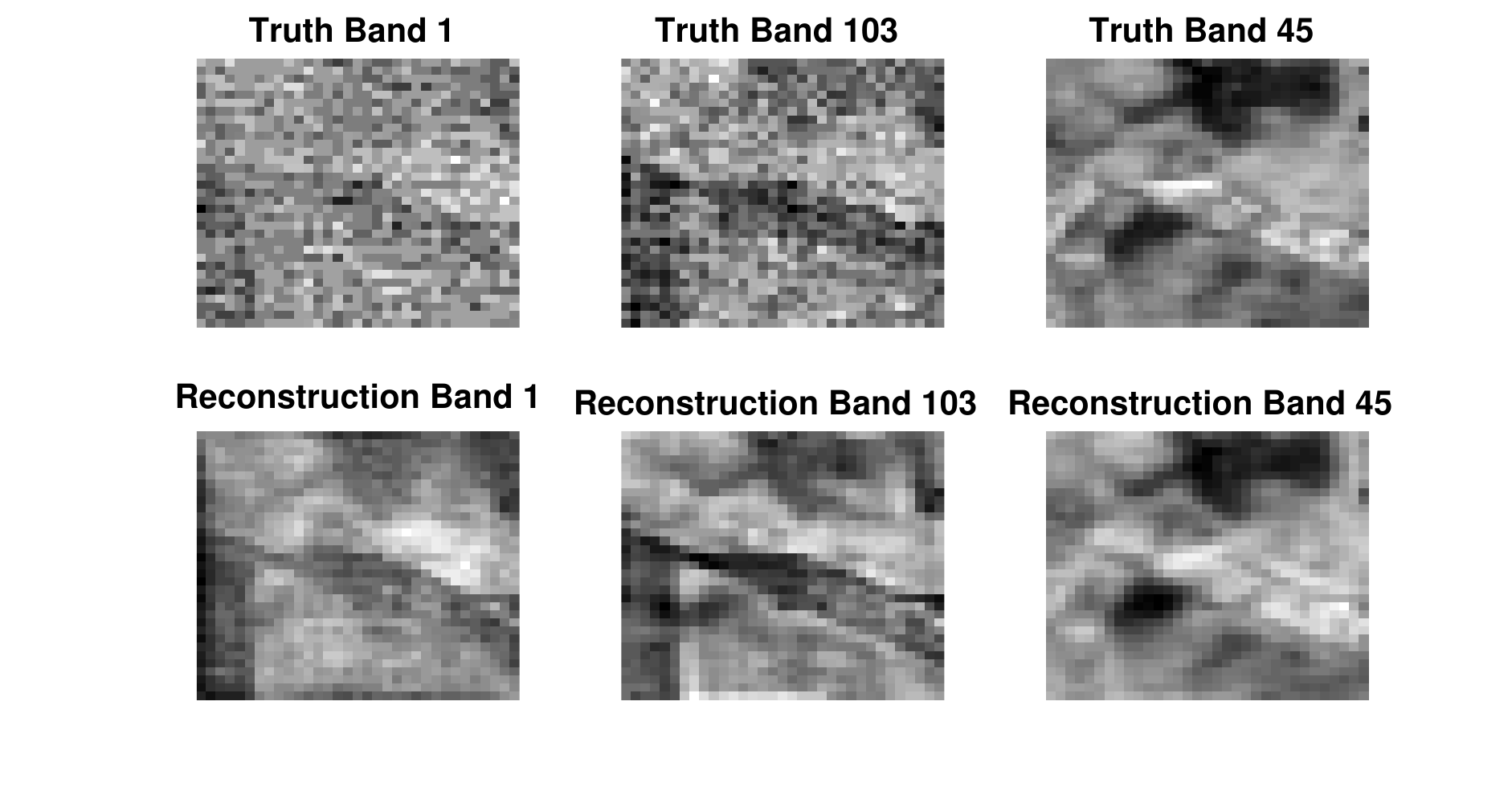}
      \end{tabular}}
  \caption{This figure shows the reconstructed and original hypercube at two noisy bands 1 \& 103 and at the clean band 45.  The reconstruction at the noiseless bands highly resemble the original image.  Although somewhat de-noised, the the images at the corrupted bands remain somewhat blurry and the presence of noise is still visible. }
  \label{fig:radonJunk}
\end{figure}

\noindent \textbf{Case III.- Simultaneous tomographic reconstruction and de-noising} -
Here we attempted to remove spectral noise from a hyperspectal data cube and reconstruct the cube from a limited number of Radon projectioLns.  In order to do so we use a 64x64 section of the original AVIRIS image as used above and observe cube through the same Radon matrix as in the flower example. The simultaneous reconstruction and spectral noise was recovered by solving the optimization problem given in (\ref{eq:minRelaxed}) the results are shown in Figure \ref{fig:radonJunk}. A $\lambda_S$ of ?? and $\lambda_L$ of ?? were used.  These reguliarization parameters were chosen using the KS-surface method described in the following section. We show good reconstruction outside of the noisy bands and significant reduction of noise within the spectrally corrupted bands.  Although the noise was somewhat reduced in the corrupted bands, as is to be expected in this limited data case, the images still remained nosier than the case when the image hypercube was directly observed with the identity operator.

\subsection{Selection of parameters $\lambda_L$ and $\lambda_S$}
\label{sect:param}
\begin{figure}[h]
\centering \makebox[0in]{
    \begin{tabular}{c}
      \includegraphics[height = 1.45in, width = 3.5in]{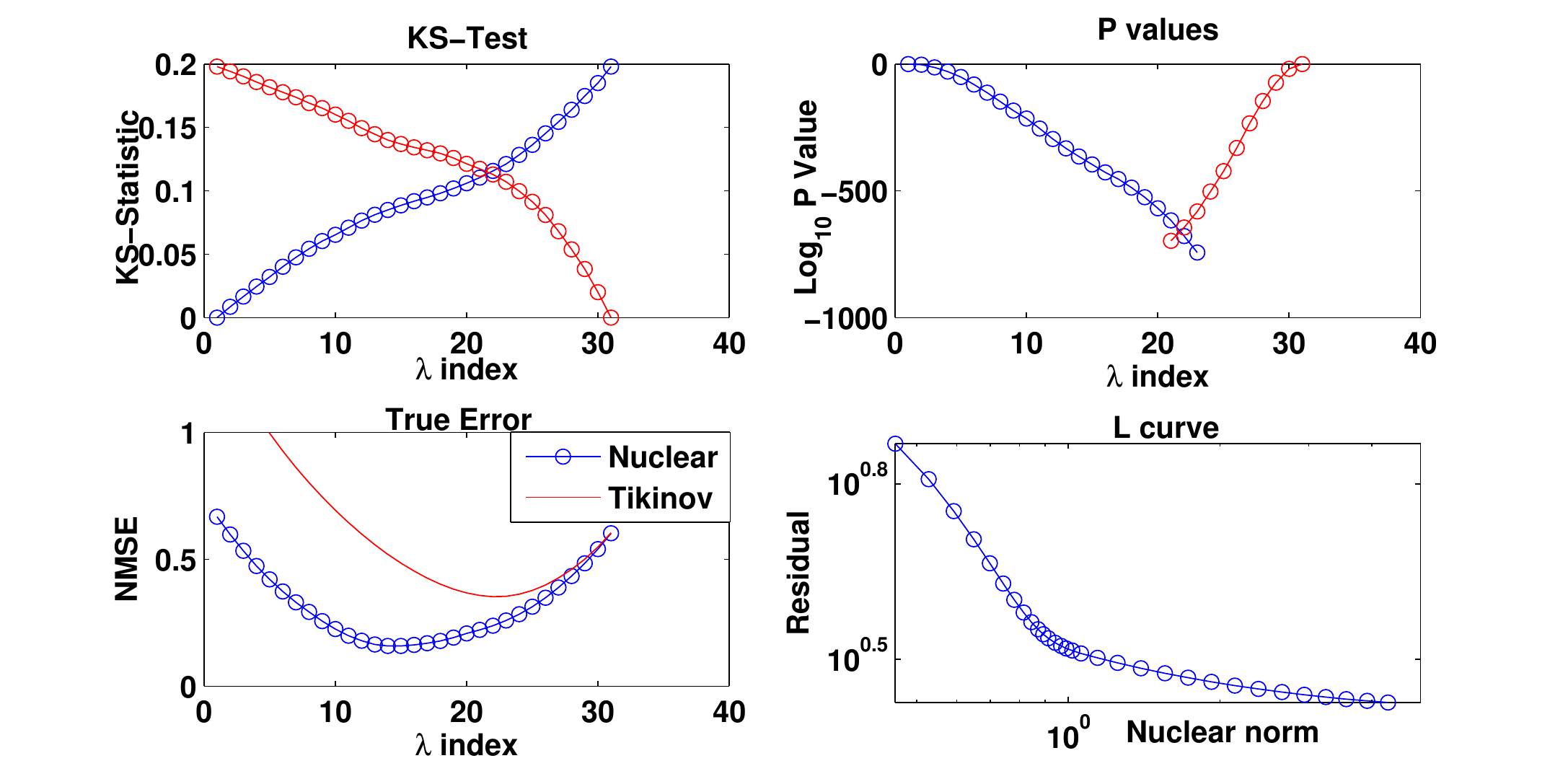}
      \end{tabular}}
  \caption{Top plots - KS test plot for recovery under limited Radon projections for the case considered. Bottom plots: (Left) - MMSE computed using the true image for various values of $\lambda$ for Tikhonov and RPCA methods; (Right) - L-curve for the RPCA method.  }
  \label{fig:ksPlot}
\end{figure}

In all of the above algorithms the issue of selection of complexity regularization parameters $\lambda$s is of practical importance. In the following experiments that we carried out on real and synthetic data sets  the choice of both $\lambda_S$ and $\lambda_L$ was determined using a one and two dimension variation of the Komolgrov-Smirnov test method proposed in \cite{AeronTSP2011}. The method essentially computes the KS test statistics of errors for a particular value of regularization parameter with respect to error residuals at extreme values of regularization parameter(s) and generates two curves. The operating point is then picked at the intersection of these two curves.\\
\textbf{Selection of regularization parameter for limited angle tomography: no spectral noise} -
The KS plot was generated with logarithmic spaced choice of $\lambda$ from $10^{-1}$ to $10^{2}$. The KS statistic values and the associated p-values are shown in the top of the Figure~\ref{fig:ksPlot} and their intersection yield a slight suboptimal $\lambda$ of 12.6. For reference the KS-test was preformed for Tikhonov regularization and the optimal, see Figure~\ref{fig:ksPlot}-bottom left plot.  From this plot we can see that for all feasible values of $\lambda$ Tikhonov regularization results in a poorer reconstruction than the nuclear norm reconstruction.  In addition to performing the KS-test for selection of regularization parameter the L-curve method \cite{Hansen1993}, commonly used for the selection of the $\lambda$ was also generated, see Figure~\ref{fig:ksPlot} bottom right.  The L-curve method results in a shallow curve without the presence of sharp knee typical of L-curve plots.  The lack of the knee makes it very difficult to select a $\lambda$ and introduces an opportunity for user bias in the selection of the regularization parameter.

\begin{figure}[h]
\centering \makebox[0in]{
    \begin{tabular}{c}
      \includegraphics[height = 1.5in, width = 3.75in]{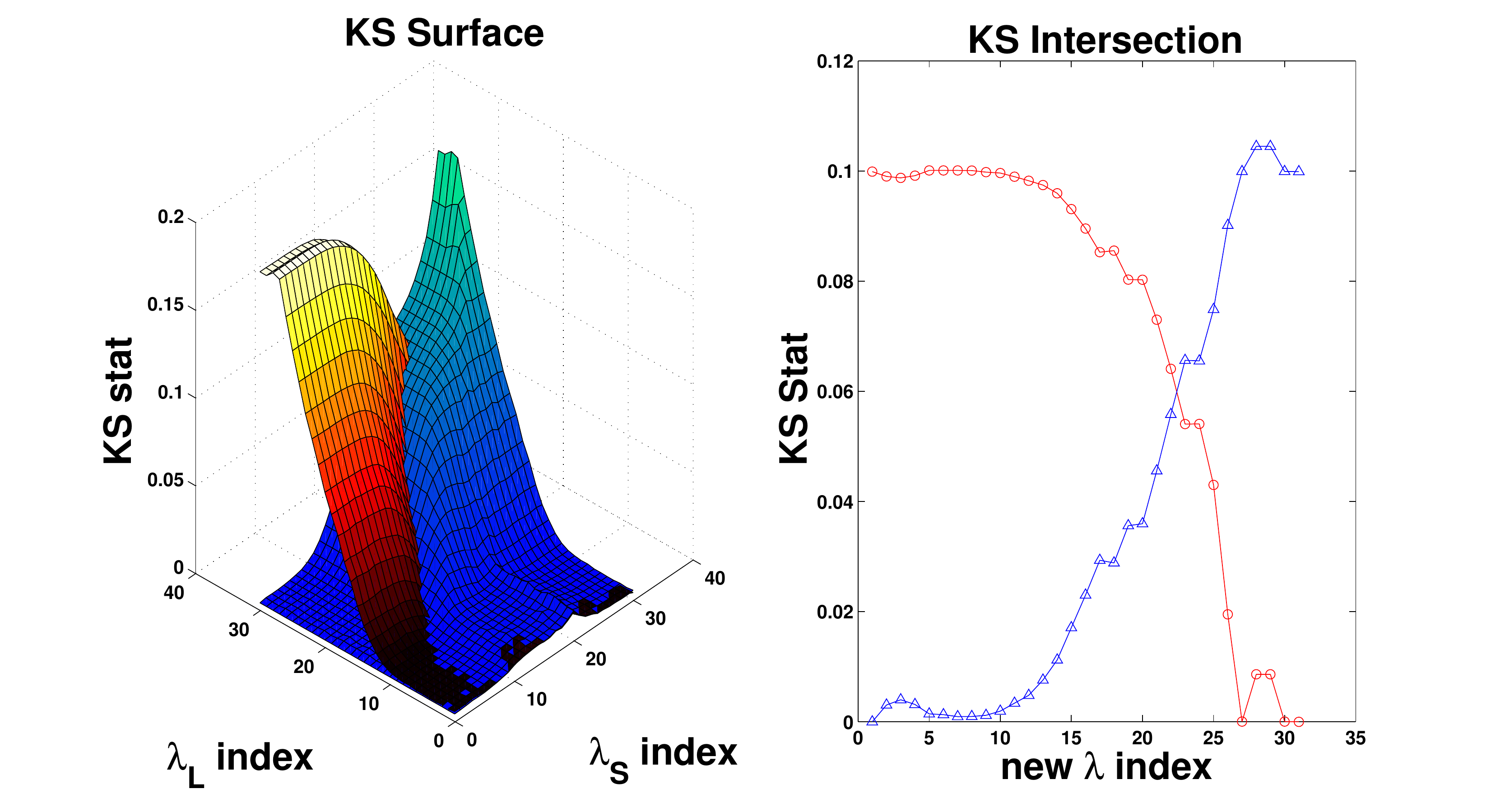}
      \end{tabular}}
  \caption{KS Surface for selecting regularization parameters for simultaneous data cube recovery and hyperspectral noise elimination. }
  \label{fig:ksSurf}
\end{figure}

\textbf{Selection of regularization parameter for limited angle tomography with spectral noise} - For this we extend the one dimensional KS test method in  \cite{AeronTSP2011} to a two dimension variation by generating a KS test surface. In order to generate the KS-surface many KS-tests were run with a fixed $\lambda_S$ and the $\lambda_L$ was varied from $10^{-3}$ to $10^{-1}$.  This process was then repeated for a range of $\lambda_S$ from $10^{-3}$ to $10^{-1}$, effectively generating a KS-plot for the selection of $\lambda_L$ for a given value of $\lambda_S$.  We can then view these multiple KS-tests as two surfaces of KS statistics as shown in Figure~\ref{fig:ksSurf} (left plot), where the intersection of the two surfaces represents best choice of $\lambda_L$ as a function of $\lambda_S$.  From this KS-surface we can then interpret the line defining the intersection of the two surfaces line of optimal regularization pairs.  We can then take the pair corresponding to the smallest value and largest value of $\lambda_L$ and generate a conventional one dimensional KS-test along the intersection of $\lambda$ pairs as shown in Figure~\ref{fig:ksSurf} right.  In this was the one dimensional KS-test selects the best pair set among the intersection of pair sets.

\newpage
\bibliographystyle{IEEEbib}
\bibliography{hyperspectral}

\end{document}